\def\mb{M_{\rm B}}
\def\mmax{M_{\rm max}}

\def\msol{M_\odot}

\def\xl{x_{l}}
\def\req{R_{\rm eq}}

\def\rms{r_{\rm ms}}
\def\mbinit{M_{\rm B,in}}
\documentclass[]{aa}
\usepackage{graphicx}
\begin{document}
\title{Spin-up of the hyperon-softened accreting neutron stars}

\author{J.L. Zdunik, P. Haensel, and M. Bejger}
\institute{
N. Copernicus Astronomical Center, Polish Academy of
Sciences, Bartycka 18, PL-00-716 Warszawa, Poland\\
e-mail: {\tt  haensel@camk.edu.pl, jlz@camk.edu.pl, bejger@camk.edu.pl}
}
\date{Received xxx Accepted xxx}
\abstract{We study the spin-up  of the accreting neutron stars
with  a realistic hyperon-softened equation of state. Using
precise 2-D calculations  we  study the evolutionary tracks of
accreting neutron stars in the
angular-momentum - frequency plane. In contrast to the case of
spinning-down solitary radio-pulsars, where a strong
back-bending behavior has been observed,  we do not see
back-bending phenomenon in the accretion-powered spinning-up
case.   We conclude that in the case of accretion-driven spin-up
the back-bending is strongly suppressed by the mass-increase
effect accompanying the angular-momentum increase.
}
\titlerunning{Back-bending in accreting NS with hyperons}

\maketitle
%
\section{Introduction}
\label{sect:introd}
%
Some  theories of dense matter predict a  phase transition to
a new ``exotic'' state  of matter at densities higher than the
nuclear saturation density. A phase transition softens the
equation of state (EOS). It has in turn been suggested that it leads to
characteristic astrophysical signatures, for example, very
large values of the braking index of spinning-down
radio-pulsars, as well  as period-clustering of spinning-up
neutron stars in low-mass X-ray binaries (Glendenning et al.
1997, Glendenning \& Weber 2001a, Glendenning \& Weber 2001b,
  Poghosyan et al. 2001, Glendenning and Weber 2002). Detection of any of these
specific features might serve as a clue to a deeper
understanding of the EOS of matter at supranuclear densities.

This work is a consecutive  step  in our studies of the
relation between the realistic EOS of dense matter, dynamics
of neutron stars, their evolutionary tracks,  and
observational signatures. As in our previous  article (Zdunik
et al. 2004, hereafter referred to as  Paper I) we would like
to focus on the phenomenon of back-bending, which is a very
characteristic consequence of the softening of the EOS  on the
spin evolution of a neutron star.

\begin{figure}[h]
\centering \resizebox{3.5in}{!}{\includegraphics{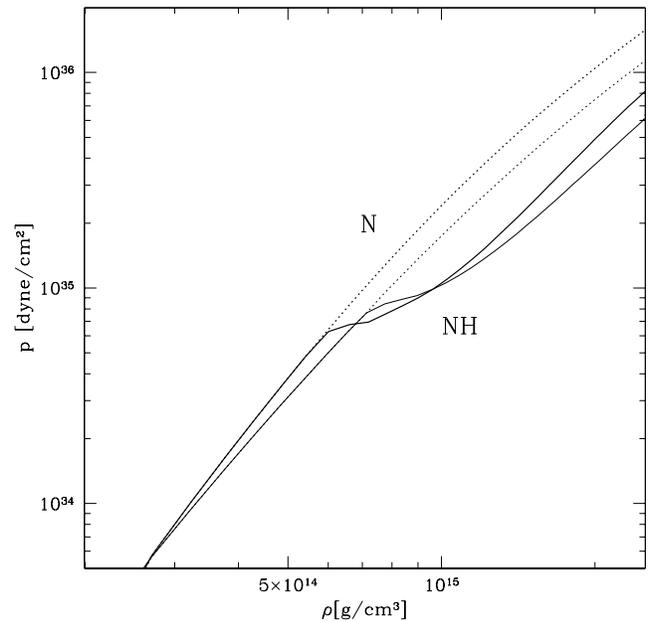}}
\caption{ Solid lines NH: pressure  $p$ vs. mass density $\rho$
for the  EoS1 $N\Lambda \Xi$ and EoS2 $N\Lambda \Xi$  of  Balberg et al. (1999),
calculated by Balberg and Gal (1997) ( N1H1 and N2H1 EOS in the notation
of Paper I). Dotted lines were  obtained by suppressing the
hyperons, i.e., by assuming purely nucleon dense matter.
 The visible softening around $\rho\sim 7\times10^{14}$
for the NH models results from the appearance of hyperons in dense
matter.} \label{fig:eos}
\end{figure}

In Paper I we showed that the back-bending in spinning-down
solitary pulsar is not necessarily evidence of an
``exotic'' phase of dense matter. Specifically, it was
demonstrated that the occurrence of back-bending can also be
caused by hyperons, particles whose  presence in dense matter
was predicted over forty years ago (Cameron 1959). We
also demonstrated the crucial importance of precision in
numerical calculations and of checking the stability of
quasi-stationary rotating configurations in the back-bending
region.

In the  present paper we describe numerical simulations of the
mass and angular momentum transfer onto a neutron star with
the EOSs  involving  hyperons. In particular, we consider the
equation EoS2 $N\Lambda \Xi$  (N2H1 in notation of Paper I) of
Balberg et al. (1999), which according to Paper I, revealed
the strongest back-bending effect in spinning-down neutron
stars.
We present results obtained for two EOSs softened by
the appearance of hyperons (out of four EOSs
 calculated  by \cite{blc99}) which
correspond to two different models of purely nucleon matter. One of
them, N1, has a standard  incompressibility of nuclear
matter $K=240$~MeV. The second model, N2, corresponds to a rather stiff
nuclear matter  with $K=320$~MeV. Consequently,
N2  gives an equation of state that is noncausal
at very high density. However, it should be stressed that
all stellar configurations presented in  our paper have
central density  well below this
non-causality threshold.

Our model  is described in Sect.\ \ref{sect:astrophys.model}.
Results of numerical calculations are collected in Sect.
\ref{sect:calcul.tracks}, while Sect. \ref{sect:concl} contains
discussion and concluding remarks.

%
\section{Astrophysical model}
\label{sect:astrophys.model}
%
We assume that the evolution of an accreting neutron star can
be represented as a sequence of stationary rotating
configurations of increasing baryon mass.  In what follows we
assume that matter is accreted from an accretion disc in which
matter elements move on Keplerian orbits, moving slowly inward
due to the viscous angular momentum outward transport.
The magnetic field is assumed to be sufficiently weak so as not to
affect the disc dynamics. However, it may still play an important
role in the angular momentum transport. The inner part of the
disc is assumed to be geometrically thin, so that the
approximation of accretion of matter in the equatorial plane
is good.

Two radii  of co-axial circles in the equatorial plane play a
crucial role in our calculation. The first one, denoted $R_{\rm
eq}$, is the radial coordinate of the neutron star surface at
the  equator. The second important radial coordinate is the
radius of the marginally stable orbit of a particle in the
space-time around and within a rotating neutron star, $r_{\rm
ms}$. Orbital motion of particles on the orbits with $r>r_{\rm
ms}$ is stable while that of particles on circular orbits with
$r<r_{\rm ms}$ is unstable. Two accretion regimes are
considered, as explained below.

 The first scenario corresponds to $r_{\rm ms}>R_{\rm eq}$.
 We then calculate the spin-up of an accreting neutron star
 according to the prescription given by Zdunik et al. (2002).
Matter is accreted from the marginally stable orbit (MSO) with
$r=r_{\rm ms}$, which determines the inner edge of the
accretion disc. A particle falls from the MSO  onto the
neutron star surface after losing an infinitesimal amount of
its angular momentum due to viscous processes. We denote
the specific angular momentum per unit baryon
mass at the MSO by $l_{\rm ms}$.
A more detailed discussion concerning determination of
MSO and $l_{\rm ms}$ in the framework of General Relativity
can be found in Zdunik et al. (2002).

In the second case $r_{\rm ms}<R_{\rm eq}$, and  MSO is located
inside the star. In this case we assume that the inner edge of
the accretion disc is connected with the stellar surface via a
thin transition layer, where viscous processes transport the
angular momentum to the neutron star.

\begin{figure}
\resizebox{\hsize}{!}{\includegraphics[]{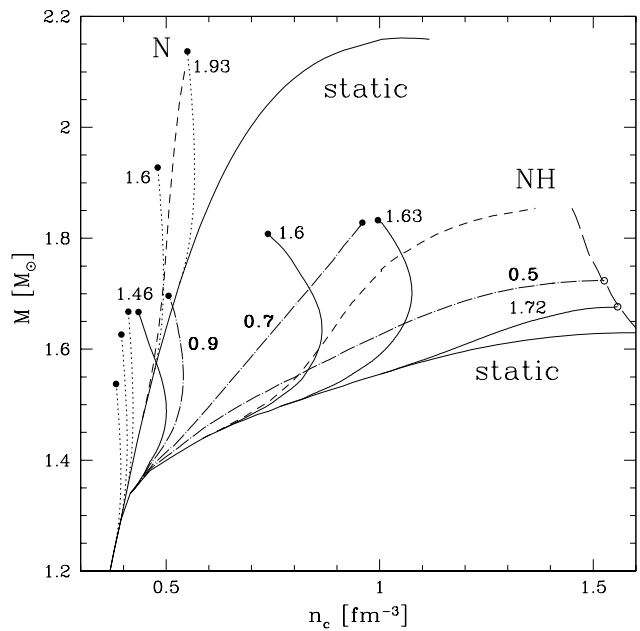}}
\caption{Gravitational mass versus central baryon number
 density along several evolutionary tracks of accreting stars
 with hyperons (model N1H1) and without hyperons (model N1).
Solid lines (model N1H1) and dotted lines (model N1)
correspond to the typical case - all angular momentum of
matter falling from the marginally stable orbit is transferred
to the star ($\xl=1$). Dash-dotted lines represent  three
cases for an $M_0=1.46~\msol$ N1H1 star with
$\xl=0.9,0.7,0.5$, fractions indicated as  boldface labels,
from the left to the right. The long-dashed, nearly vertical line
at maximum mass is defined by the onset of instability with
respect to axi-symmetric perturbations. Dashed lines correspond
to the condition $\req=\rms$.
Termination points are marked by a filled
circle for Keplerian configurations and by an open circle for
the instability with respect to axi-symmetric perturbation. }
 \label{fig:nm}
\end{figure}

In both cases we define the innermost stable circular orbit
(ISCO), from which the angular momentum and baryon mass is
transferred to a neutron star. The ISCO determines the inner
edge of the accretion disc. We use the approximation $r_{\rm
ISCO}=r_{\rm ms}$ for $r_{\rm ms}>R_{\rm eq}$ and $r_{\rm
ISCO}=R_{\rm eq}$ for $r_{\rm ms}<R_{\rm eq}$.

The value of specific angular momentum per unit baryon mass
of a particle orbiting the neutron star at the ISCO, $l_{\rm IS}$,
can be calculated by solving exact equations of the orbital
motion of a particle in the space-time produced by a rotating
neutron star, given in Appendix A of Zdunik et al.
(2002).

\begin{figure}[h]
\centering \resizebox{3.5in}{!}{\includegraphics{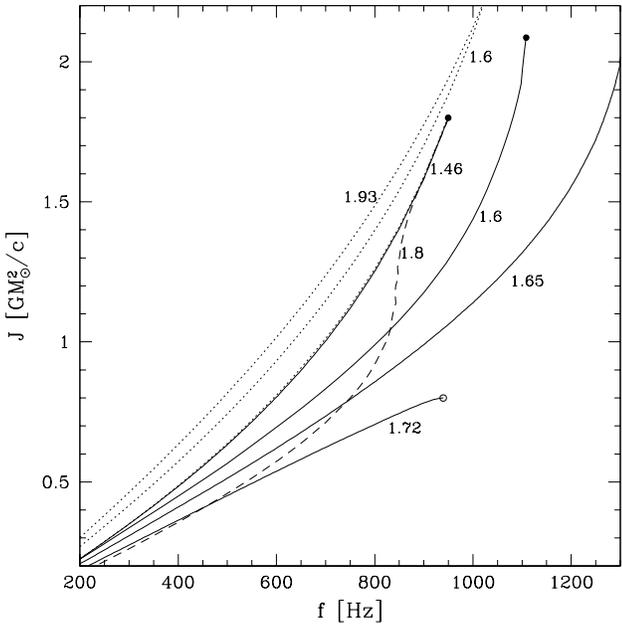}}
\caption{ Angular momentum as a function of spin frequency for
accreting, spinning-up NS, with $x_l=1$ for EOS with (solid
lines, N1H1 model) and without hyperons (dotted lines, N1 model).
The label gives
the baryon mass for the starting non-rotating
configuration.  The $J(f)$ track of an isolated NS with baryon
mass of $1.8~M_\odot$ (thin dashed line) is also shown. To
make comparison easier,  the initial rotating state of
spinning down isolated NS has been chosen to lie on the
spin-up track of accreting NS with initial baryon mass of
$1.46~M_\odot$.
 Termination points are marked by a filled circle
for Keplerian configurations and by an open circle for the
instability with respect to axi-symmetric perturbation.
}
\label{fig:fj_cmp}
\end{figure}

Accretion of an infinitesimal amount of baryon mass ${\rm
d}M_{\rm B}$ onto a rotating neutron star is assumed to lead
to a new quasi stationary rigidly rotating configuration of
mass $M_{\rm B}+{\rm d}M_{\rm B}$ and angular momentum $J+{\rm
d}J$, with
\begin{equation}
{\rm d}J= x_{l}l_{\rm IS} \;{\rm d}M_{\rm B}~, \label{eq:lms}
\end{equation}
where  $x_l$ denotes the fraction of the angular momentum of
the matter element transferred to the star. The remaining
fraction $1-x_l$ might be lost via radiation or  other
dissipative  processes. In the case of $r_{\rm ms}>R_{\rm eq}$,
these processes are  acting on the MSO and during the fall of
a matter element  through the dilute but magnetized plasma and
the photon radiation emitted from stellar surface, which fills
the gap between the MSO and stellar surface (Agol \& Krolik
2000, Afshordi \& Paczy{\'n}ski 2003).  In the case of $r_{\rm
ISCO}=R_{\rm eq}$, the dissipation and the angular momentum
loss take place within the \textit{boundary  layer}, through
which the accretion disk joins the stellar surface (see, e.g.,
Popham \& Sunyaev 2001, and references therein).
%
\section{Calculations of  evolutionary tracks}
\label{sect:calcul.tracks}
%
We assume that the evolution of an accretiting neutron star can
be represented as a sequence of stationary rotating 2-D
configurations.  The numerical method used for the calculation
of configurations belonging to this  family was described
in Sect.\ 3 of Paper I. Here, we would like to stress the very
high precision of our calculations; evaluation of the GRV2 and
GRV3 virial error indicators (see Nozawa et al. 1998 and
references therein) showed relative errors within $\simeq
10^{-5}$.

\begin{figure}
\resizebox{\hsize}{!}{\includegraphics[]{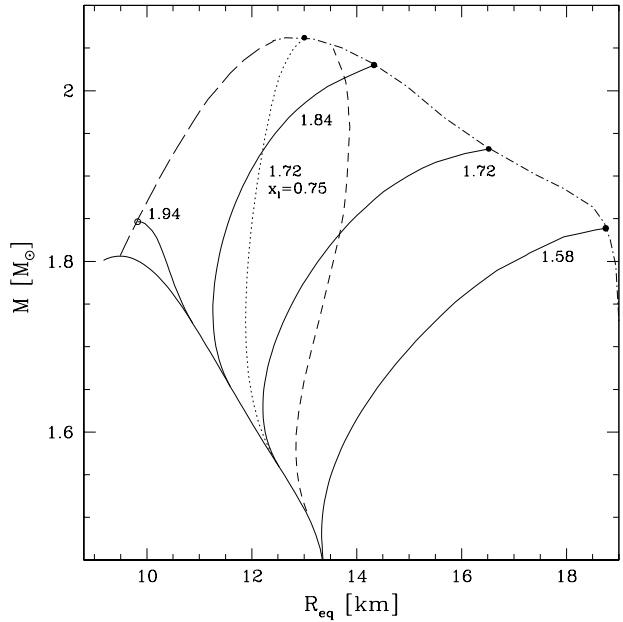}}
\caption{Gravitational mass versus the (circumferential) equatorial radius,
along  several evolutionary tracks of accreting strange stars
for N2H1 model.
 The dotted line represents  the model for which only 75\% of angular momentum is
 transferred to the star. The long-dashed line at
maximum mass is defined by the onset of instability with
respect to axi-symmetric perturbations. The dash-dot line
corresponds to the Keplerian (mass shedding) limit. The dashed
line at $R_{\rm eq}\simeq 13$~km corresponds to the condition
$\req=\rms$. }
 \label{fig:rm}
\end{figure}
%
\begin{figure}[h]
\centering \resizebox{3.5in}{!}{\includegraphics{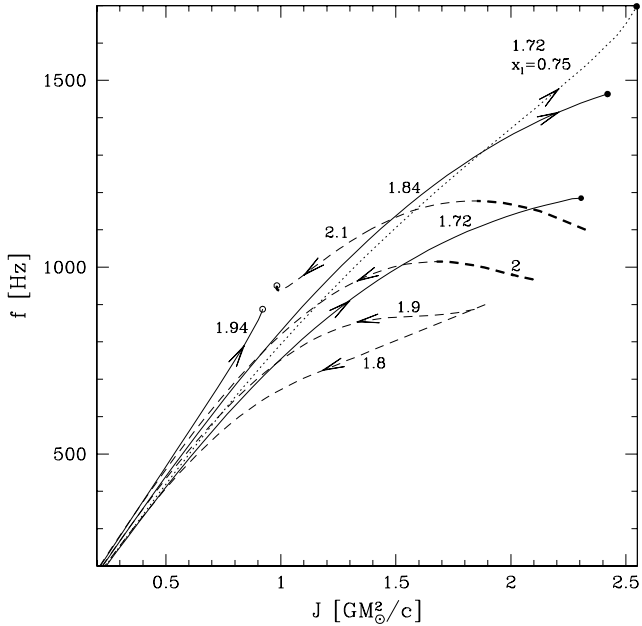}}
\caption{Spin frequency as a function of angular momentum of
the solitary, spinning-down pulsars (dashed lines) and
accreting, spinning-up NS with $x_l=1$ (solid lines). In the
case of dashed lines, the numbers denote their
baryon mass $M_{\rm B}$ in the units of $M_\odot$, which
stays constant along the evolutionary track. For accreting
neutron-star spin-up tracks, the label gives the baryon
mass for the starting non-rotating configuration. The arrows
indicate the evolution of neutron-star configurations with
time.  Termination points are marked by a filled circle for
Keplerian configurations and by an open circle for the
instability with respect to axi-symmetric perturbation. For the
$2.1M_\odot$ curve, notice  a tiny thick segment,
corresponding to the relativistic spin up by the angular
momentum loss, just before the axi-symmetric instability limit
is reached.  Note the change of sign of ${\rm d}f/{\rm d}J$
for the dashed (isolated pulsars) curves, which signals the end
point of the back-bending stage, denoted by thick dashed line
segments. There is no back-bending for accreting stars of
similar baryon masses, for which ${\rm d}f/{\rm d}J$ stays
positive along all evolutionary track. For comparison we
plotted one curve with $x_l=0.75$ (dotted line).}
\label{fig:fj}
\end{figure}

To study the effect of hyperons on the evolution of accreting
neutron star, we compare the results for N1 and N1H1 EOSs, in
the notation of Paper I,  models of Balberg \& Gal (1997). In Fig.
\ref{fig:nm} we present the evolutionary tracks in $n_{\rm
c}-M$ plane, where $n_{\rm c}$ is the central baryon density
and $M$ is gravitational mass of NS. Using this figure we can
answer the question how accretion changes physical conditions
in the center of a star. As we can see for $\xl=1$, the curves
for N (nucleons only) stars are almost vertical. It means that
accretion results in the increase of $f$ (and $J$), but the
density in the center of the star remains essentially
unchanged.  In the case of NH EOS (with hyperons), at  the
first stage of accretion the density in the center increases,
as well as  the size of the hyperon core. However, the main
effect of the mass and angular momentum transfer is spin-up
and  change of shape (deformation) of NS, accompanied by a
decrease of $n_{\rm c}$. As in the case of strange stars
(Zdunik et al. 2002), the effect strongly depends on the
fraction, $x_l$,   of the angular momentum transferred to the
stars from the ISCO. For $\xl=0.5$, central density and the
size of hyperon core increase during accretion,  leading
eventually to a collapse of the NS into a black hole; this is
to be contrasted with the $\xl>0.7$ cases, where the
termination point is the mass-shedding (Keplerian) limit.
Considered examples show that, in order to build a rather
large hyperonic core, a significant  amount of angular
momentum has to  be lost in the course of accretion.

Another  difference between the evolution of accreting NS with
and without hyperon core is  visualized in Fig.\
\ref{fig:fj_cmp}. For the N1 and N1H1 EOSs, the evolution of
spin frequency, as a function of angular momentum,  depends on
the mass of initial configuration in opposite ways. For
$\mbinit=1.46\msol$,  the evolution for these two EOSs does
not differ significantly, and the curves almost coincide. The N1
stars with higher initial mass have larger $J$ for the same
$f$. If a star contains a hyperon core, as for the N1H1 model,
higher mass results in a lower $J$ for a given spin frequency.
This effect can be explained by the specific dependence of the
moment of inertia $I$ on the mass of the star. For slow
rotation, the moment of inertia of the star is defined as a
limit $\displaystyle{I=\lim_{f \to 0} {\rm d}J/{\rm d} f}$.
Let us recall that, except for $M$ close to maximum allowable
mass for non-rotating NSs, $M^{\rm stat}_{\rm max}$, the value
of $I$ increases monotonically with $M$. However, close to
$M^{\rm stat}_{\rm max}$, a rapid decrease of $R$ with
increasing $M$ dominates and makes $I$ decrease with
increasing $M$.  For the N1 model the initial configuration with
$\mb \sim 1.5 - 2 \msol$ is sufficiently  far from the maximum
mass and the increase with $M$ dominates. For the N1H1 model we are
close to $\mmax$ and the  decrease of $R$ with a relatively small
change (increase) in $M$ dominates.

 In Fig. \ref{fig:fj_cmp} we also plot the sequence of
 spinning down configurations with hyperon core
  for isolated neutron star
(fixed baryon mass $\mb$). To make comparison easier, the
baryon mass for this sequence ($1.8~\msol$) is equal to the
final mass of the accreting model with the initial mass
$1.46~\msol$. The difference between the $J(f)$ dependence is
evident; for isolated NS during slowing down the pulsar
loses about half of its initial angular momentum without
changing its rotation period much (in this case $\sim
800~{Hz}$). This effect is almost completely suppressed by
accretion, as the derivative ${\rm d}J/{\rm  d}f$ does not
change significantly along the curve $M_{\rm B, in}=1.46\msol$

The same effect can be seen for the N2H1 EOS, for which we
obtained the most pronounced back-bending behavior for the
isolated, spinning down neutron stars (Paper I). Results are
presented in figures \ref{fig:rm} and \ref{fig:fj}. In
Fig.~\ref{fig:rm} we also plot the curve defined by the
condition $\req=\rms$, which is almost vertical and corresponds
to the equatorial radius 13~km. Marginally stable orbit exists
to the left of this curve (i.e. equatorial radius smaller than
$\sim 13$~km). The evolution of accreting NS, when  75\% of
the angular momentum is transferred to the star,  can lead to
the maximum accreted mass larger by more than $0.1\msol$ than
in the case of $x_l=1$ (typically the mass increase is $\sim
0.4\msol$), before the star reaches the Keplerian limit (see
Fig. \ref{fig:rm}). In Fig. \ref{fig:fj} we use angular
momentum as an argument (abscissa), because the change of angular
momentum is directly connected with a time evolution, by Eq.\
(\ref{eq:lms}),  in the case of accretion and by the equation
describing the energy and angular momentum loss due to the
physical processes (e.g. magnetic dipole radiation) for
isolated spinning  down NS. The value of $l_{\rm IS}$ does not
change significantly during accretion period,  so that  the
change of $J$ is roughly proportional to the accreted mass
(i.e., to  $\dot M\cdot t$).

For isolated neutron stars regions of back-bending exist
(in Fig. \ref{fig:fj} for $2.1\msol$) and are denoted by thick dashed
lines. The tiny  thick-line region of spin-up of the star by
angular momentum loss just before the onset of the
axi-symmetric instability (open circle), is a direct
consequence of stability condition in GR (for other examples
of this phenomenon, see Figs.\ 7-9 in Zdunik et al. 2004).

 As one sees,
there is no back-bending phenomenon for the accreting NSs,
even in the most favorable case $\xl=1$, where the spinning-up
of NS is the most efficient. This is due to a basic difference
in the spin evolution of isolated and accreting NSs. For
accreting NSs, increase in $J$ is necessarily accompanied with
an increase in $M_{\rm B}$. This clearly weakens
the impact of rotation on  the stellar structure because of
a simultaneous increase in $f$ and in the gravitational
pull. Therefore, in this case spin-up by accretion cannot influence the
structure of rotating NS so strongly as to enforce the
back-bending behavior.

The total baryon mass accreted during the accretion
spin-up epoch is obviously limited by the mass of the
companion in the LMXB. This results in an upper bound on the
 final frequency reached at the end of accretion.
It may happen, for example, that the spin-up tracks
in Fig. 3 terminate well before the instability points.

%
\section{Conclusion}
\label{sect:concl}
We studied the spin-up of accreting NSs with hyperonic cores
and, in particular, the impact of the hyperon softening of the
EOS of dense matter on the spin-up history. The spin-up
sequences ended either at the Keplerian (mass shedding) limit
or, for sufficiently high initial baryon mass, by an
instability with respect to axi-symmetric perturbations
leading to collapse into a Kerr black-hole. We considered an
EOS that exhibits a strong back-bending behavior in the spin-down
of isolated NSs. However, this EOS does not show back-bending
in the spin-up by accretion, even in the most efficient
spinning-up, without any angular momentum loss in the accretion
process. Therefore, for hyperonic EOSs back-bending in the
pulsar timing does not imply spin clustering in accreting NSs
in low-mass X-ray binaries (LMXBs). Generally, our
calculations show that additional increase of mass
that accompanies increase of angular momentum in accreting NSs,
weakens or even suppresses the back-bending phenomenon in the
time evolution of NS rotation frequency.
It should be stressed
that this suppression of the BB phenomenon for hyperon EOSs does not mean that
the effect cannot be theoretically obtained for very
strong (perhaps unrealistically strong) softening of the EOS.
In such a case, the   back-bending present in the
spin-down timing will  be also present (albeit
in a much weaker form)  in the accretion
spin-up, provided, of course, that not too large fraction of the angular
momentum is lost during the accretion process.
This aspect of the
impact of the high-density softening of the EOS on the spin
evolution of neutron stars and, in particular, its
observational signatures in the timing of isolated and
accreting NSs will be the subject of our forthcoming
publication.
\acknowledgements{This work was partially supported by the
Polish MNiI Grant No. 1 P03D 008 27 and by the CNRS/PAN LEA
Astr-PF.  }
%

\end{document}